# Monolithic growth of ultra-thin Ge nanowires on Si(001)

J. J. Zhang,[1,]* G. Katsaros,[1,§] F. Montalenti,[2] D. Scopece,[2] R. O. Rezaev,[1,4] C. Mickel,[1] B. Rellinghaus,[1] L. Miglio,[2] S. De Franceschi,[3] A. Rastelli,[1,§] and O. G. Schmidt[1]

[1]*IFW Dresden, Helmholtzstr. 20, D-01069 Dresden, Germany*
[2]*L-NESS and Materials Science Department, University of Milano-Bicocca, I-20125 Milano, Italy*
[3]*SPSMS, CEA-INAC/UJF-Grenoble 1, F-38054 Grenoble Cedex 9, France*
[4]*Laboratory of Mathematical Physics, Tomsk Polytechnic University, Lenin Ave. 30, 634050 Tomsk, Russia*

Self-assembled Ge wires with a height of only 3 unit cells and a length of up to 2 micrometers were grown on Si(001) by means of a catalyst-free method based on molecular beam epitaxy. The wires grow horizontally along either the [100] or the [010] direction. On atomically flat surfaces, they exhibit a highly uniform, triangular cross section. A simple thermodynamic model accounts for the existence of a preferential base width for longitudinal expansion, in quantitative agreement with the experimental findings. Despite the absence of intentional doping, first transistor-type devices made from single wires show low-resistive electrical contacts and single hole transport at sub-Kelvin temperatures. In view of their exceptionally small and self-defined cross section, these Ge wires hold promise for the realization of hole systems with exotic properties and provide a new development route for silicon-based nanoelectronics.

* E-mail: j.zhang@ifw-dresden.de
§ Present address: Institute of Semiconductor and Solid State Physics, University Linz, A-4040 Linz, Austria

As miniaturization in complementary metal oxide semiconductor transistors proceeds and approaches the atomic scale, the reliability and reproducibility of transistors become increasingly difficult because of random fluctuations in the number of dopants included in the active device. Furthermore, as dimensions shrink, surface defects present in top-down etched structures become more and more detrimental.

In this context, Ge-based semiconducting nanowires (NWs) are attracting great interest [1-3]. Doping-free Ge/Si core/shell NWs with diameters of 20 nm were used to fabricate field effect transistors that showed performances comparable to state-of-the-art devices fabricated by conventional lithographic top-down processes [2, 4]. Outstanding electrical properties, such as ballistic conduction up to length scales of several hundred nanometers, were reported in such core/shell NWs [5]. Recently, atomic-scale NWs could be fabricated on Si(001) and Ge(001) surfaces using a lithography technique based on scanning tunneling microscopy and a gaseous dopant source [6, 7]. Such wires were created by direct dopant implantation and Ohm's law was observed to hold at the atomic scale [7], making them suitable as interconnects. While miniaturization poses problems for applications, it opens up many possibilities for investigating fundamental physics. Indeed transport through single dopants has been observed [8, 9] and the realization of spin qubits has become possible [10]. A very recent proposal [11] has further suggested that *ultra-thin, strained* Ge NWs can support helical modes, which renders them appealing for realizing spin filters [12], Cooper-pair splitters [13] and for observing exotic quantum states, like Majorana fermions [14-17].

Ge NWs are commonly obtained by vapour-liquid-solid growth, in which a metallic catalyst nanoparticle initiates and sustains the growth of a wire out of the substrate plane [1]. The use of metallic catalysts, however, introduces metal contamination [18], making the integration with microelectronics technology rather problematic. It is also a formidable challenge to transfer and arrange these vertically grown NWs to an adequate substrate for device fabrication.

Alternatively, already in 1993, Tersoff and Tromp [19] suggested that catalyst-free, ultra-thin Ge "quantum wires" with large built-in strains could be grown epitaxially on flat Si substrates. In this letter, we show that such types of wires can indeed be obtained by a self-assembled process implemented in a solid-source molecular beam epitaxy (MBE) growth setup. The self-assembly of the Ge NWs is achieved through a surprisingly simple procedure consisting of the epitaxial deposition of a Ge layer on a Si(001) substrate followed by thermal annealing at appropriate temperatures. Compared to NWs grown by catalytic methods [1-3], the catalyst-free Ge NWs presented here exhibit an outstanding uniformity in their lateral size, they lie horizontally along well-defined crystallographic directions (either [100] or [010]), and they are monolithically integrated into the silicon substrate. Our theoretical calculations show that the formation of {105} facets plays a key role in determining the stability and uniformity of the wires. The successful realization of good electrical contacts to individual wires and the observation of single-hole transport make them a promising system for realizing both ultra-small p-type Ge transistors on Si and novel quantum devices.

The Ge NWs were grown by MBE at a base pressure of $5\times10^{-11}$ mbar. We initially deposit 4.4 monolayers (ML) of Ge to form a pseudomorphically-strained two-dimensional layer, known as the wetting layer (WL), with a growth rate of 0.04 ML/s at a substrate temperature of 570°C. The deposited Ge amount is slightly smaller than the critical thickness of 4.5 ML for the formation of three-dimensional (3D) Ge islands, referred to as "hut clusters" [20-25]. After Ge deposition, the substrate temperature is kept at nominal 560°C for different time durations. During this *in-situ* annealing, 3D islands appear and evolve into long wires via anisotropic growth along either the [001] or the [010] crystallographic direction, as shown in Fig. 1(a). This finding indicates that, for the chosen amount of deposited Ge, the WL is metastable against 3D island formation [26]. The length of the wires is typically already a few hundreds of nanometers after 1 h annealing and reaches the micrometer scale in 3 h. Further annealing produces only a limited increase in their length, as shown in Fig. 2(a). This may be attributed to the fact that, as Ge moves into the wires, the WL is consumed leading to gradual reduction of the growth rate due to the depletion of the Ge supersaturation [24]. As seen from Fig. 1(a) the wires are highly uniform in height and width. A statistical analysis performed on NWs longer than 80 nm shows an average height *h* of 1.86 nm (about three unit cells) with a remarkably low standard deviation (0.14 nm). The NWs have the triangular cross-section characteristic of hut clusters (Fig. 1(d)), with {105} side facets forming an angle $\theta = 11.3°$ with the substrate plane and resulting in an average base width $b = 2h/\tan\theta = 18.6$ nm. This is confirmed by cross-sectional transmission electron microscopy (TEM) as shown in the inset of Fig. 1(d). Figure 2(b) shows the histograms of height distribution of all nanostructures including short hut clusters and pyramids (as seen in Fig. 1(a)) after 1, 3, and 12 h annealing. Different from the length, the height distribution does not show significant variations during annealing, indicating the wires grow by increasing only their length [24]. In addition, pyramids and hut clusters usually have a larger height (> 2 nm), compared to the wires, suggesting that islands with a large height are difficult to elongate.

The wire density can be controlled simply by the amount of the initially deposited Ge. By decreasing it, the amount of metastable Ge is correspondingly decreased, resulting in a reduced island nucleation rate [27]. Under these conditions the wire density drops (Fig. 1(b)), but the maximum wire length increases up to $L \sim 2$ μm, which corresponds to a length $L$ to height $h$ ratio as large as ~1000. By increasing the initial Ge amount, a larger density of comparatively shorter wires is obtained (see supplemental material [28]). We attribute this observation to two factors: (i) an initially larger island density leads to an increased probability of "collisions" between growing wires and consequent interruption of wire growth due to strain repulsion [29]; (ii) the Ge material available for each wire decreases, so that even relatively isolated wires cannot grow too long.

We find that the Ge wires have a constant height (width) as long as they grow on the same atomic terrace. When their length extends over several terraces, we observe the top ridge of the wires to remain parallel to the (001) plane, at least for moderate local miscut angles. In other words, as a wire crosses an atomic step on the underlying substrate its height will increase or decrease by 0.14 nm (i.e. the height of an atomic step). This implies that by choosing the morphology of the Si surface prior to growth, the size of the wires can be tuned at the atomic scale. On substrates with larger terraces, which may be obtained as in Ref. [30], we expect all the wires to have a constant height (width). On substrates with smaller terraces, tapered wires are instead observed (Fig. 1(c)).

The Ge wires do not consist of pure Ge due to the Si-Ge intermixing taking place during Ge deposition and the subsequent annealing process. Although the ultra-small dimension of the wires does not allow us a more detailed determination of the composition, our selective wet chemical etching in $H_2O_2$ solution shows that the Ge content is higher than 65% even at the base which is known to have the lowest Ge content [28].

Let us now show that there exists a strong thermodynamic driving force stabilizing long {105}-faceted wires. We evaluate the energy difference $\Delta E$ between a wire on an N-layer thick WL, and a configuration where the same material is instead spread on the WL, creating a region with N+1 layers. Here all parameters are quantified by considering pure Ge. Due to the large aspect ratio $L/b$ of interest, the energy change can be computed by disregarding wire terminations (see inset of Fig.3, and discussion in the Supplemental Material [28]). In this way we consider directly mature huts neglecting the first seeds and their initial stages of growth and elongation. Since this processes are likely to involve atomic-scale effects [21-25], further investigations are needed to capture their physics and evolution, but this is out of the scope of the present work. By taking into account elastic-energy relaxation, surface energy differences, and wire edge energies, simple calculations lead to [28]:

$$\Delta E = V\left[\Delta\rho_{eff} + \frac{4}{b\tan\theta}\Delta\gamma + \frac{4\Gamma}{b^2\tan\theta}\right] \quad (1),$$

where $V$ is the wire volume, $\Gamma$ the total energy associated with edges connecting adjacent facets, and $\Delta\rho_{eff} = \Delta\rho_{el} + (1/h_1)[\gamma_{WL}(N) - \gamma_{WL}(N+1)]$, i.e. the sum of the elastic energy (per unit volume) lowering, provided by the wire geometry and the surface-energy change obtained by adding the (N+1)$^{th}$ layer (with height $h_1$) to the WL [28, 31]. Finally, $\Delta\gamma = \gamma_{hut}\sec\theta - \gamma_{WL}(N)$, where $\gamma_{hut}$ is the surface energy (per unit area) of the wire. $\Delta\rho_{el}$ was quantified by finite element method (FEM) calculations, surface energies were extracted from *ab-initio* calculations [28, 32], while $\Gamma$=370 meV/Å accounts for wire edge energies, as obtained from experimental fitting in Ref. [33]. To mimic the experiment, we set N=4. A plot of $\Delta E/V$ vs. $b$, shown in Fig. 3 demonstrates that: (i) There exists a "magic" base width $b_{min} \approx 2\Gamma_{hut}/-\Delta\gamma$, minimizing the energy of a wire at fixed volume, and (ii) Wire formation is energetically favourable vs. WL thickening, as indicated by the corresponding negative value of $\Delta E$. These results are a direct consequence of the very low surface energy of Ge{105} under compressive strain [33, 34], $\Delta\gamma \approx -4.5$ meV/Å$^2$, implying that the driving force for wire formation is the reduction of surface energy, rather than strain relaxation (which is one order of magnitude smaller [28]). This energy gain is however counterbalanced by the edge energy, dominating at small $b$ values, and resulting in a favoured width. Remarkably, the theoretical estimate

$b_{min} \approx 16.3$ nm is within about 15% of the experimentally observed value. The presence of a $V$-independent minimum in the $\Delta E/V$ curve, explains not only the sharp distribution of NW cross-sectional sizes but also the tendency towards "infinite" elongation. In addition, one notices an asymmetric behavior in terms of $\Delta E/V$ vs. $b$ around the minimum. For $b<b_{min}$, the quantity $\Delta E/V$ increases much faster than for $b>b_{min}$. This leads to the prediction that it is energetically easier to increase the cross-section of a wire rather than decrease it. This is exactly what we observe in our experiments (Fig. 1): The smaller ends of tapered wires have a rather uniform width of about 16 nm, which is very close ($\Delta b/b_{min}$~15%) to the experimentally determined $b_{min}$ (18.6 nm). In contrast, the larger ends have a broad width distribution extending up to 40 nm (Fig. 1(c)), corresponding to ($\Delta b/b_{min}$~115%). We therefore believe that on stepped terraces the wires preferentially grow on their larger ends with lower surface and keep growing in the opposite direction of tapering. We conclude that the model provides an excellent explanation of the main experimental findings.

After over two decades of research on the Ge/Si epitaxial system, it may seem surprising that the nanowire growth method presented above has not been reported before. In fact, although conceptually simple, this method requires certain growth conditions to be met. For instance, any initially crowded environment (in terms of critical nuclei) would not allow the observation of micron-long wires, because of self-blocking (due to strain repulsion) and/or coarsening. With this respect, the annealing of an initially flat WL with proper thickness seems to be a key to reduce the density of mobile species leading to clustering. A too large amount of Ge or too high temperature during growth or subsequent annealing would again increase such density (the thicker the WL, the weaker the atomic bonds [31]). On the other extreme, too low temperature and/or too thin WL would simply suppress both wire nucleation and elongation through surface diffusion.

In view of their extremely small and uniform cross section, the Ge NWs reported here are excellent candidates for the realization of novel electronic devices. To this aim, a new set of samples were grown in which the Ge NWs were covered by a few-nm-thick Si cap layer to create core-shell structures. The Si cap layer was grown at a relatively low temperature of 300°C in order to reduce intermixing and obtain a sharp Si/Ge interface, as seen from the TEM image in Fig. 1(d). FEM calculations show that the Ge NWs are partially strained without the Si cap [28] and become almost fully strained to the Si lattice in the growth plane with a 2 nm-thick Si cap (Fig. 4(a)). Three-terminal, field-effect devices were fabricated out of individually contacted core-shell wires (Fig. 4(b)) [28]. The metallic contacts were deposited close to each other, defining a 30-50 nm wide channel. At room temperature, these devices are shunted by a significant leakage current through the Si substrate. Therefore, their basic electrical properties were only studied at low temperature using a $^3$He refrigerator. Figure 4(c) shows a representative measurement of the source-drain current ($I$) as a function of the top-gate voltage ($V_{TG}$) at 260 mK. The $I(V_{TG})$ dependence confirms the p-type character of the NWs, originating from the type II band alignment between Si and Ge and from the pinning of the contact Fermi level near the Ge valence-band edge. The device can be tuned from a fully pinched-off state for $V_{TG} > 0$, to a relatively low-resistance state (40 kΩ) for $V_{TG} < 0$. Remarkably, $I$ as high as a few μA, corresponding to current densities of $10^7$ A/cm$^2$, could be driven through the Ge NW. At small source-drain voltage ($V_{SD}$), the $I(V_{TG})$ characteristic exhibits a sequence of narrow peaks as shown in the inset of Fig. 4(c). From a 2D plot of $|I|$ as a function of $V_{TG}$ and $V_{SD}$, shown in Fig. 4(d), we ascribe these peaks to single-hole transport occurring at the degeneracy between the consecutive charge states of a *single* quantum dot, created between the source and drain metal contacts. In each of the diamond-shape regions, transport is blocked by a Coulomb energy barrier and the quantum dot holds a well-defined, integer number of holes.

The above results lay the ground for a range of fundamental studies and device applications at low temperature. The operation of Ge-NW devices may be further extended to room temperature by replacing the Si substrate with silicon on insulator (SOI) substrates with a very thin Si surface layer. In this perspective we have successfully grown Ge NWs on SOI substrates with a 35-nm-thick Si surface layer [28], paving thus the way towards the realization of devices operating at room temperature.

We acknowledge the financial support by the DFG SPP1386, P. Chen and D. J. Thurmer for MBE assistance, R. Wacquez for providing the ultrathin SOI wafers, and G. Bauer, Y. Hu, X. Jehl, S. Kiravittaya, C. Klöffel, E. J. H. Lee, F. Liu, D. Loss, S. Mahapatra for helpful discussions. G. K. acknowledges support from the European commission via a Marie Curie Carrer Integration Grant. S.D.F. acknowledges support from the European Research Council through the starting grant program.

[1] A. M. Morales and C. M. Lieber, Science **279**, 208 (1998).

[2] J. Xiang *et al.*, Nature **441**, 489 (2006).

[3] S. Kodambaka, J. Tersoff, M. C. Reuter, and F. M. Ross, Science **316**, 729 (2007).

[4] N. Singh *et al.*, IEEE Trans. Electron Device **55**, 3107 (2008).

[5] W. Lu, J. Xiang, B. P. Timko, Y. Wu, and C. M. Lieber, Proc. Natl. Acad. Sci. **102**, 10046 (2005).

[6] G. Scappucci *et al.*, Nano Lett. **11**, 2272 (2011).

[7] B. Weber *et al.*, Science **335**, 64 (2012).

[8] M. Pierre *et al.*, Nat. Nanotechnol. **5**, 133 (2010).

[9] M. Fuechsle *et al.*, Nat. Nanotechnol. **7**, 242 (2012).

[10] Y. Hu, F. Kuemmeth, C. M. Lieber, and C. M. Marcus, Nat. Nanotechnol. **7**, 47 (2012).

[11] C. Kloeffel, M. Trif, and D. Loss, Phys. Rev. B **84**, 195314 (2011).

[12] P. Streda and P. Seba, Phys. Rev. Lett. **90**, 256601 (2003).

[13] L. Hofstetter, S. Csonka, J. Nygard, and C. Schönenberger, Nature **461**, 960 (2009).

[14] J. Alicea, Phys. Rev. B **81**,125318 (2010).

[15] R. M. Lutchyn, J. D. Sau, and S. Das Sarma, Phys. Rev. Lett. **105**, 077001 (2010).

[16] Y. Oreg, G. Refael, and F. von Oppen, Phys. Rev. Lett. **105**, 177002 (2010).

[17] V. Mourik, K. Zuo, S. M. Frolov, S. R. Plissard, E. P. A. M. Bakkers, and L. P. Kouwenhoven, Science **336**, 1003 (2012).

[18] J. E. Allen *et al.*, Nat. Nanotechnol. **3**, 168 (2008).

[19] J. Tersoff and R. M. Tromp, Phys. Rev. Lett. **70**, 2782 (1993).

[20] Y. W. Mo, D. E. Savage, B. S. Swartzentruber, and M. G. Lagally, Phys. Rev. Lett. **65**, 1020 (1990).

[21] D. E. Jesson, G. Chen, K. M. Chen, and S. J. Pennycook, Phys. Rev. Lett. **80,** 5156 (1998).

[22] M. Kästner and B. Voigtländer, Phys. Rev. Lett. **82,** 2745 (1999).

[23] I. Goldfarb, L. Banks-Sills, and R. Eliasi, Phys. Rev. Lett. **97,** 206101(2006).

[24] M. R. McKay, J. A. Venables, and J. Drucker, Phys. Rev. Lett. **101**, 216104 (2008)

[25] L. V. Arapkina and V. A. Yuryev, J. Appl. Phys. **109**, 104319 (2011)

[26] B. J. Spencer, P. W. Voorhees, and S. H. Davis, J. Appl. Phys. **73**, 4955 (1993).

[27] J. Tersoff and F. K. LeGoues, Phys. Rev. Lett. **72**, 3570 (1994).

[28] See supplemental material for a detailed description of the materials and methods of both experiments and theoretical calculations.

[29] H. T. Johnson and L. B. Freund, J. Appl. Phys. **81**, 6081 (1997).

[30] S. Tanaka, C. C. Umbach, J. M. Blakely, R. M. Tromp, and M. Mankos, Appl. Phys. Lett. **69**, 1235 (1996).

[31] M. Brehm *et al.*, Phys. Rev. B **80**, 205321 (2009).

[32] D. Scopece, F. Montalenti, and M. J. Beck, Phys. Rev. B **85**, 085312 (2012).

[33] G. Chen *et al.*, Phys. Rev. Lett. **108**, 055503 (2012).

[34] G. H. Lu, M. Cuma, and F. Liu, Phys. Rev. B **72**, 125415 (2005).

Fig. 1 (color online). Atomic force microscopy (AFM) images of Ge wires forming on Si(001) substrates after 12 h annealing. Atomic terraces are parallel to the (001) plane and atomic steps on the WL are well visible. (a) High and (b) low density of Ge wires on Si(001) with a nominal miscut angle of less than 0.05°. (c) Tapered Ge wires on Si(001) with a nominal miscut angle of less than 0.5°. The wires grow laterally along either of the two <100> directions as indicated by the arrows and their surface is composed of four {105} facets. Scale bar: 200 nm. (d) 3D AFM image of an individual Ge wire. The inset is a cross-sectional TEM image of the Ge wire capped with Si at 300°C, showing a sharp Si/Ge interface and an inclination angle of 11.3° between {105} facets and the substrate plane. Scale bar: 5 nm.

Fig. 2. Histograms showing the length distribution (a) and the height distribution (b) of Ge nanostructures (including wires, pyramids and hut clusters) for different annealing times at a substrate temperature of 560°C.

Fig. 3 (color online). The energy difference $\Delta E$ (divided by volume) between a wire and a 2D configuration of equal $V$, as obtained using Eq. (1), is plotted vs. base width $b$. The inset illustrates the structures used in the model: truncated wires with only two {105} facets. Points along the curve in the plot represent wires of different length but same volume, as sketched in the inset. The black filled circle indicates the base width $b_{min}$, corresponding to the minimum-energy configuration. $b_{min}$ is volume independent.

Fig. 4 (color online). (a) Schematic of a Ge wire capped with a 2 nm-thick Si cap. The in-plane (left) and out-of-plane (right) components of the strain distribution are shown. (b) Schematic of a device showing the Ge wire contacted by Al electrodes (gray), covered with a ~10 nm hafnia layer (blue) and a layer of Ti/Pt (10/90 nm) acting as a top gate (green). The top left scheme shows a cross-section along the wire. (c) $I$ vs. $V_{TG}$ at $V_{SD}$=75 mV. The device can be switched off at about 600 mV while currents higher than 1 µA can flow through the wire at high negative gate voltages. For $V_{SD}$=0.25 mV (inset), characteristic peaks originating from Coulomb blockade can be observed. (d) $|I|$ vs. $V_{TG}$ and $V_{SD}$, revealing Coulomb diamonds and charging energies as high as 25 meV. The conductance throughout the plot is reduced at zero bias due to the superconducting properties of the Al electrodes. This can be seen more clearly in a second device [28].

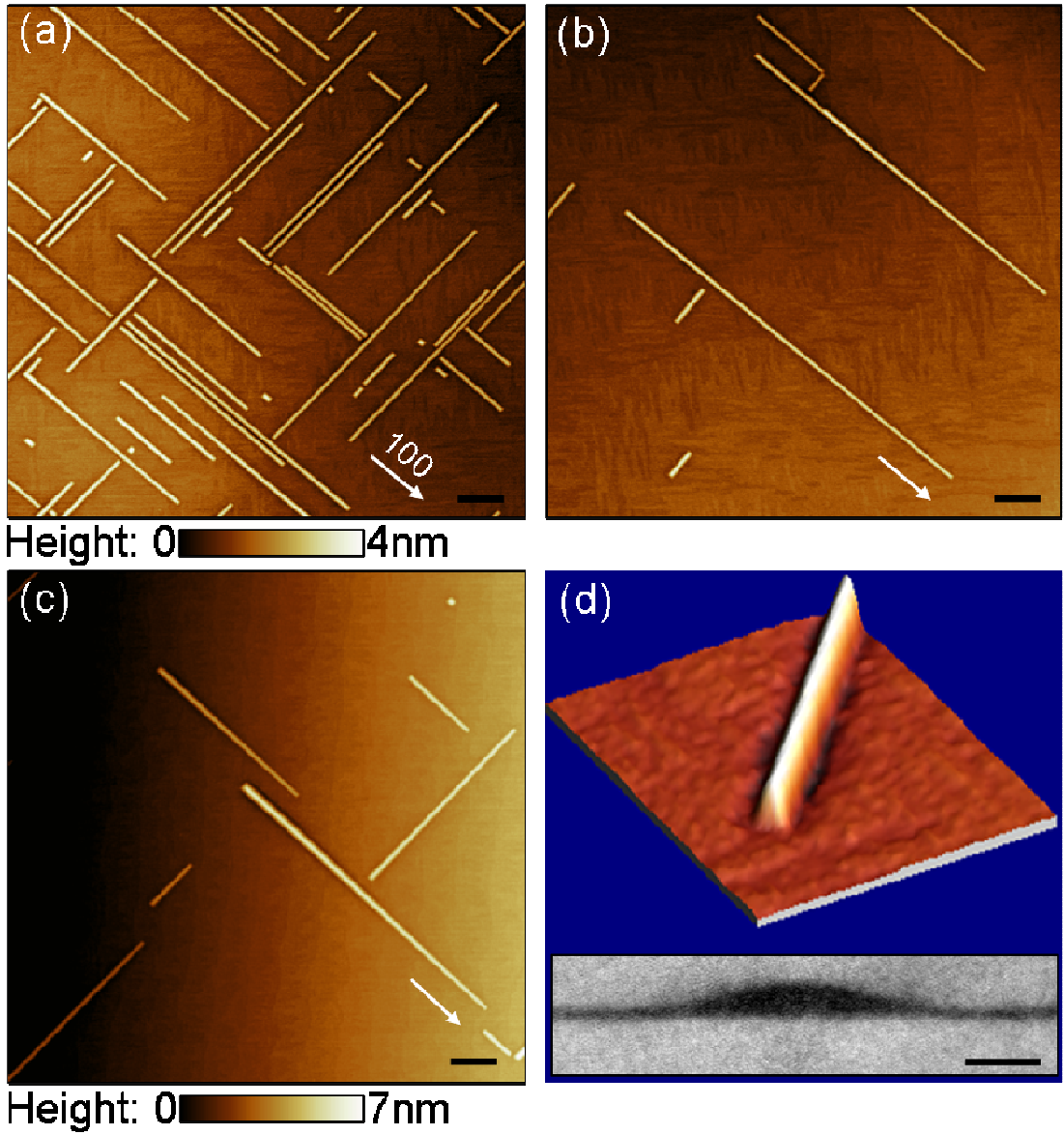

**Figure 1**

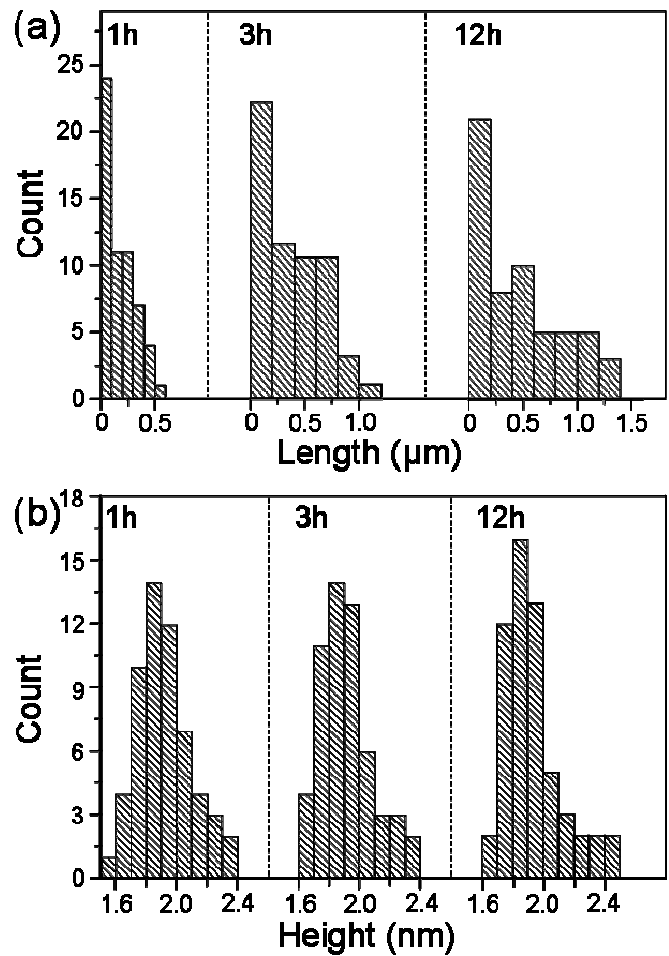

Figure 2

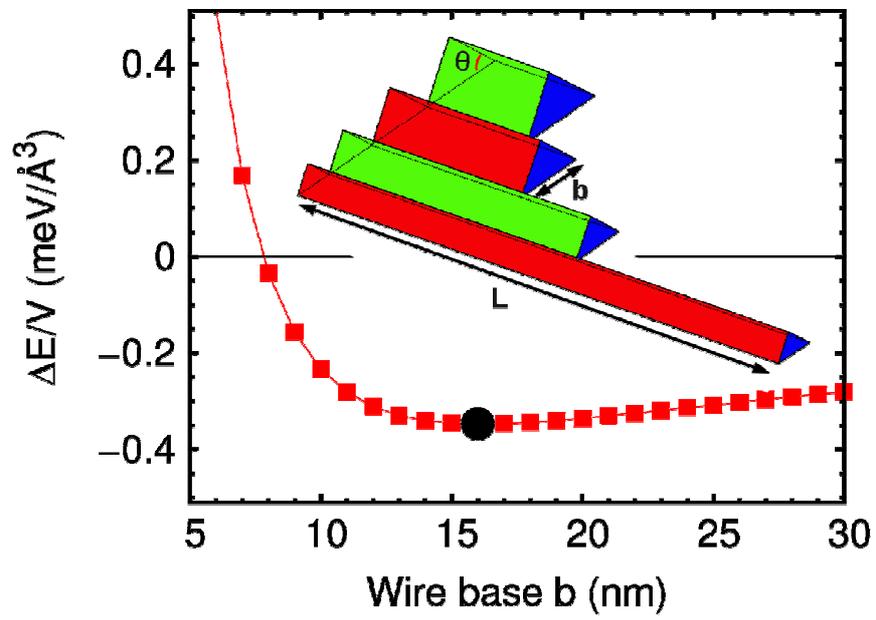

**Figure 3**

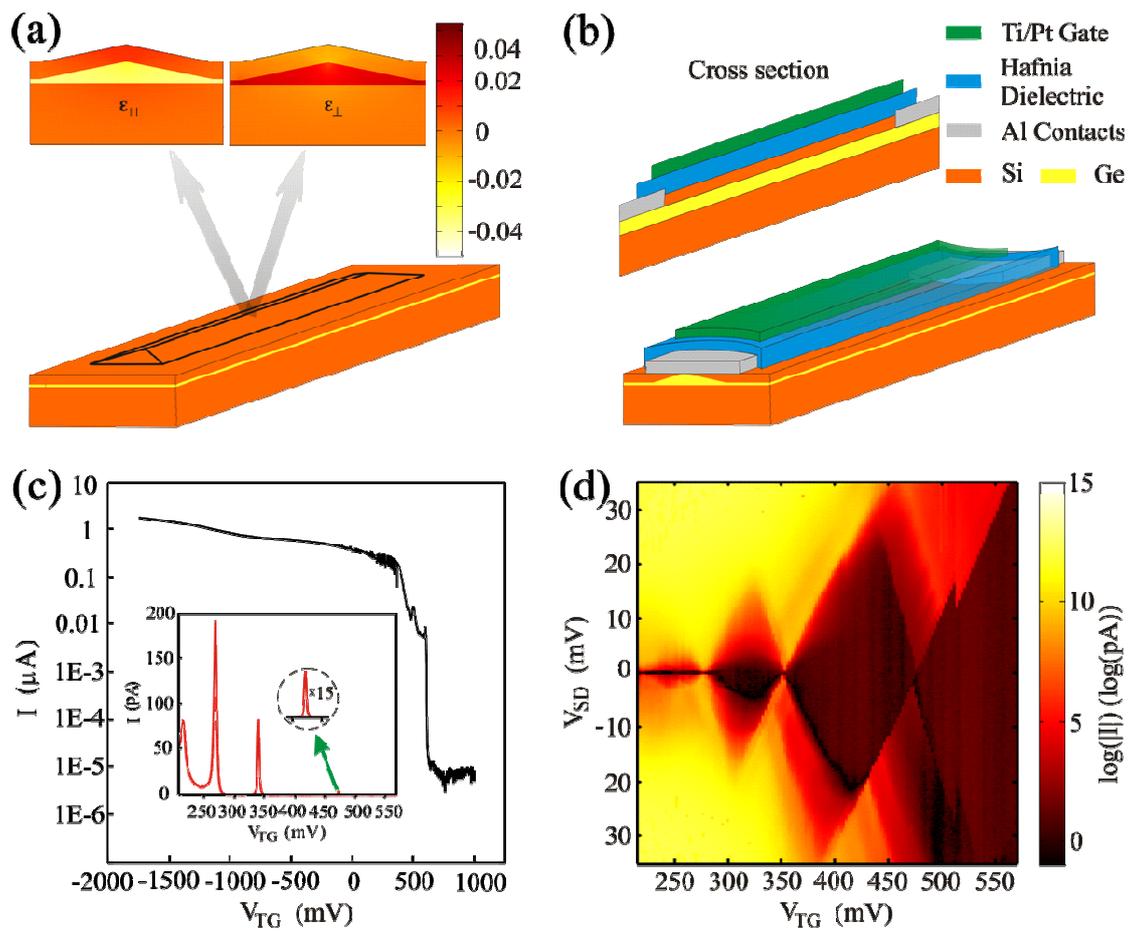

**Figure 4**

# Monolithic growth of ultra-thin Ge nanowires on Si(001)

J. J. Zhang,[1,*] G. Katsaros,[1,§] F. Montalenti,[2] D. Scopece,[2] R. O. Rezaev,[1,4] C. Mickel,[1] B. Rellinghaus,[1] L. Miglio,[2] S. De Franceschi,[3] A. Rastelli,[1,§] and O. G. Schmidt[1]

[1]*IFW Dresden, Helmholtzstr. 20, D-01069 Dresden, Germany.*

[2]*L-NESS and Materials Science Department, University of Milano-Bicocca, I-20125 Milano, Italy.*

[3]*SPSMS, CEA-INAC/UJF-Grenoble 1, F-38054 Grenoble Cedex 9, France.*

[4]*Laboratory of Mathematical Physics, Tomsk Polytechnic University, Lenin Ave. 30, 634050 Tomsk, Russia*

* E-mail: j.zhang@ifw-dresden.de.

§ Present address: Institute of Semiconductor and Solid State Physics, University Linz, A-4040 Linz, Austria

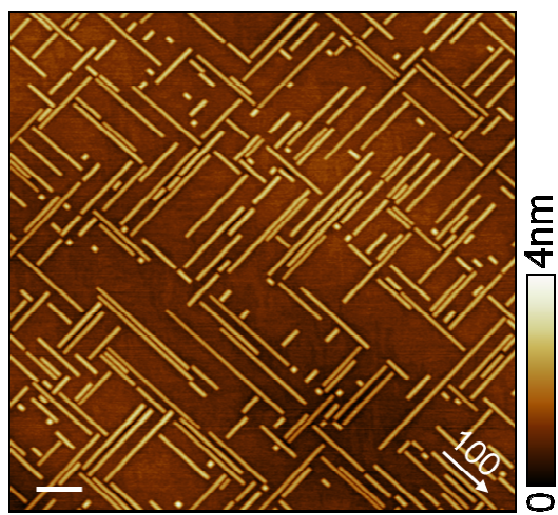

Fig. S1. AFM image of Ge wires with a larger density of comparatively shorter length. The amount of initially deposited Ge is 4.6 ML, where sparse hut clusters are formed. Scale bar: 200 nm.

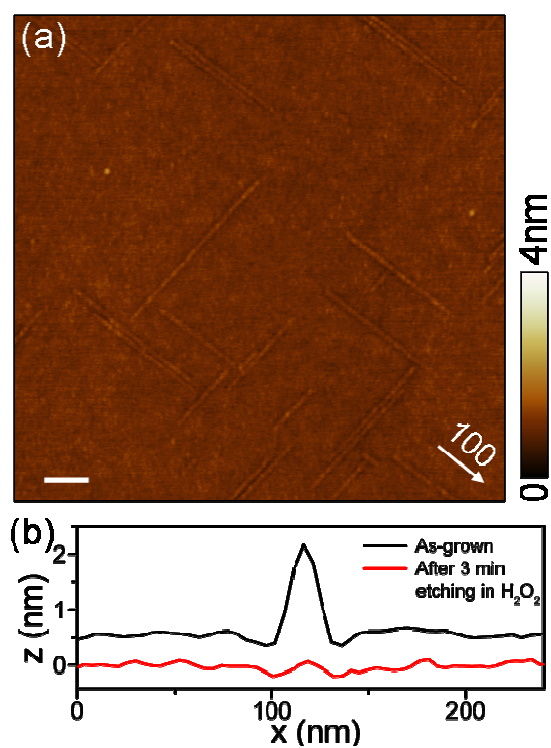

Fig. S2. (a) AFM image of Ge wires after 3 min etching in 31% $H_2O_2$ solution. Scale bar: 200 nm. The $H_2O_2$ solution selectively etches $Si_{1-x}Ge_x$ with $x$ larger than ~ 0.65 [1]. From the image it can be seen that after 3 minutes they have been etched, indicating that the Ge content is higher than 65%. (b) Height profiles taken perpendicularly across the wires before (black) and after (red) selective etching in the $H_2O_2$ solution. The offset between them corresponds to the thickness of the etched wetting layer.

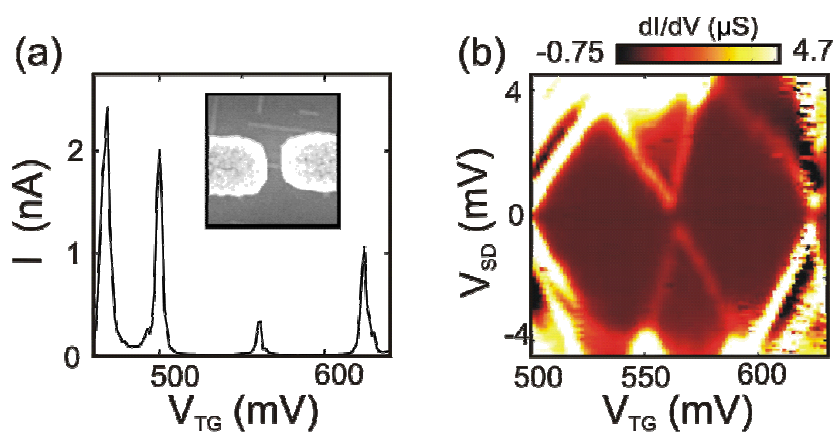

Fig. S3. (a) $I$ vs. $V_{TG}$ at $V_{SD}$=1mV for a second device. The inset shows a SEM image of a typical device before the fabrication of the top gate. (b) d$I$/d$V$ vs. $V_{SD}$ and $V_{TG}$ for the same

device. The diamonds are not closed because of the superconducting gap present in the Al electrodes.

## Growth substrates:

The Si wafers on which the Ge wires have been grown and transport measurements have been performed are intrinsic (n-) wafers with a resistance higher than 3000 Ohm*cm. For the growth on silicon on insulator (SOI) wafers, substrates with a buried oxide of 145 nm and a Si device layer of 21 nm were used.

## Device Fabrication:

Standard optical lithography was used in order to define the bonding pads and the alignment crosses needed for electron beam lithography. In a following step 2×2 µm² metallic (Cr/Au 2/8 nm) squares with a distance of 5 µm were defined all over the write field. 200 nm wide and 30 nm thick Al electrodes, with a gap of about 30-50 nm separating them, were overlaid on top of these metallic layers. This was done in order to achieve good contacts to the Al in the following step. By means of scanning electron microscopy those pairs of Al electrodes contacting a single wire were identified and Cr/Au contacts were used to connect the Al electrodes to bonding pads. In a next step a 2×2 µm² window was opened on top of the wires and 100 cycles of hafnia were deposited at 130°C by means of atomic layer deposition. Following lift-off and another step of electron beam lithography a Ti/Pt 10/90 nm metallic film, used as a top gate electrode, was deposited by electron beam evaporation.

## Theory: model parameters and wire stability

In our manuscript we use a simple model to establish the relative stability of high aspect-ratio NWs (or huts) of different dimensions. Here we briefly outline it and we describe how the various parameters were computed and/or inferred from existing literatures.

Following Ref. [2], we compare the energy of a wire (or hut) of volume $V$ placed on a N-layer thick wetting layer (WL) with the one of a configuration where the material stored in the wire is re-distributed to create a portion of an additional monolayer (ML), of height $h_1$. The two configurations are sketched in Fig. S4(a). By calling $S_{hut}$ the

{105} surfaces exposed by the wire, $B_{hut}$ the WL surface covered by it, and $\lambda_{hut}$ the total length of the edges, the energy difference between the two configurations is:

$$\Delta E = V\left[\Delta\rho_{el} + \frac{\gamma_{WL}(N) - \gamma_{WL}(N+1)}{h_1}\right] + S_{hut}\gamma_{hut}(\varepsilon, N) - B_{hut}\gamma_{WL}(N) + \lambda_{hut}\Gamma \quad (S1)$$

where $\Delta\rho_{el}$ is the difference between the elastic energy (per unit volume) of the configurations reported in Fig. S4(a). The surface energies $\gamma's$ depend in general on the strain state $\varepsilon$ of the surface and on the distance between the free surfaces from the Ge/Si interface [3]. This distance is conveniently expressed in number N of MLs (Fig. S4(a)). A simple exponential form [3] is used to describe the full dependence on N:

$$\gamma(N) = (\gamma_0 - \gamma_\infty) \cdot e^{-BN} + \gamma_\infty \quad (S2)$$

where $\gamma_0$ and $\gamma_\infty$ are the surface energies of the pure Si (N=0) and pure Ge surface, respectively, and $B$ is a parameter in unit of ML$^{-1}$. The (001) surface (i.e. the free surface of the WL) is uniformly subject to biaxial compressive strain since it adapts to the lattice of the Si substrate. Instead, every point of the exposed {105} facets of the wires relax differently, making the dependence on strain non trivial. Additionally, every point at the tilted surface is at a different distance from the Si/Ge interface. The multiscale method we employ to handle this case is the one described in Ref. [3], the only differences being that FEM calculations of the strain field have been here repeated for Ge huts on Si(001), and that the (1 1 10) WL is here replaced by a (001). In our model, we employ surface energy values as determined from DFT-LDA and, in order to guarantee internal consistency, the elastic energy densities $\rho$ are determined from a FEM calculation where the elastic constants are the ones predicted by *ab-initio* calculations [2, 4]. In Eq. (S1), $\Gamma$ is the total energy per unit length of the edges, whose determination is out of reach for *ab-initio* calculations. In Ref. [4], however, a comparison between theory and experiments for ripples on Si(1 1 10) leads to an estimate of 370 meV/Å. The observed disordered boundary between crossing {105} facets at the ripples' top offers an explanation for the high value. As STM images of huts on Si(001) [5, 6] also displayed similar features, we tentatively used the exact same value (assigning it solely to the top, disordered edge), despite edges in huts and ripples connect differently oriented {105} facets. With this choice the results nicely agree with the experiments, without requiring ad-hoc fitting, as reported in the main manuscript.

The geometrical parameters describing a hut of aspect ratio $r = L/b$ in Eq. (S1) are:

$$V = \tfrac{1}{12} \cdot b^3 \cdot \tan\theta \cdot (3r-1) \qquad (S3)$$

$$S_{hut} = b^2 \cdot r \cdot \sec\theta \qquad (S4)$$

$$B_{hut} = b^2 \cdot r \qquad (S5)$$

$$\lambda_{hut} = b \cdot \left[ r - 1 + 2\sqrt{1 + \sec^2\theta} \right] \qquad (S6)$$

where $\theta = 11.3°$.

Since the wires observed experimentally have a high length-to-base ratio we can simplify Eqs. (S3)-(S6) in the limit of $r \gg 1$. This is equivalent to considering the energy of a vertically cut portion of the wire as sketched in Fig. S4(b), where the energies of both the vertical cutting surfaces and the relative edges are neglected. For L/b~50 (still shorter than the observed wires) the neglected fraction of volume, surface and edge amount to ~ 1%, 2%, 5% of the complete hut respectively, so that our approximation is expected to nicely hold. Notice that the simplified geometry also allows for simpler two-dimensional FEM calculations of the strain tensor in the hut, needed to estimate both $\gamma_{hut}$ and $\Delta\rho_{el}$.

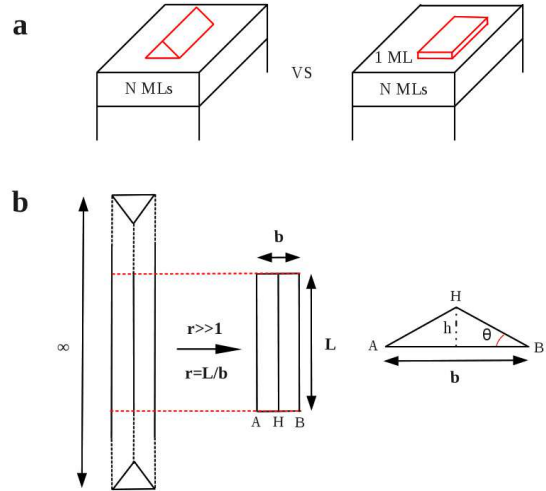

Fig. S4: (a) Eq.(S1) yields the energy difference between a wire laying on a N-layers thick WL (left), and a configuration where the atoms in the wire re-arrange into a flat configuration, forming a partial (N+1)[th] layer (right), bounded by monoatomic steps whose contribution to the total energy is neglected. The elastic energy density of the configuration on the right is taken to be the one of a perfect WL, therefore neglecting lateral relaxation. (b) Sketch of the geometrical simplification, adapted to model wires with large length-to-base aspect ratio $r$. $\theta = 11.3°$ is the inclination angle between the exposed {105} facets and the substrate (001) plane.

Within this approximation, Eq. (S1) can be expressed as:

$$\Delta E = V \cdot \left[ \Delta \rho_{eff} + \frac{4}{b \tan \theta} \Delta \gamma + \frac{4\Gamma}{b^2 \tan \theta} \right] \quad (S7)$$

Where $\Delta \rho_{eff} = \Delta \rho_{el} + [\gamma_{WL}(N) - \gamma_{WL}(N+1)]/h_1$ and $\Delta \gamma = \sec \theta \cdot \gamma_{hut}(\varepsilon, N) - \gamma_{WL}(N)$.

Exploiting Eq. (S7), one can readily estimate the preferred base size $b_{min}$, among huts of assigned volume. For the experimentally-relevant case N=4, numerical minimization with the above specified parameters yields $b_{min} \approx 16$ nm. A full plot of $\Delta E/V$ vs. $b$ for N=4 is reported in Fig. 3 of the main manuscript.

Hidden in Eq. (S7), but included in the numerical treatment, are the variations of both $\Delta \rho_{el}$ and $\Delta \gamma$ with $b$. The first, determined by the compressive lobes caused by the hut in the WL [7] is extremely weak in the general case (see Fig. S5(a) for the case of N=4 ML and $b$ =16 nm). The second, caused by the variation with $b$ of the local distance between the different portions of the {105} facets and the Si interface, becomes negligible only for thick WL, i.e. when the exponential in Eq. (S2) is close enough to its limiting value (this is the case for N=4 as Fig. S5(b) shows). This would allow us to estimate $b_{min}$ from Eq. (S7) analytically:

$$b_{min} = \frac{2\Gamma}{-\Delta \gamma} \quad (S8)$$

For N=4, the estimate provided by Eq. (S8) is very good, yielding $b_{min} \approx 15$nm.

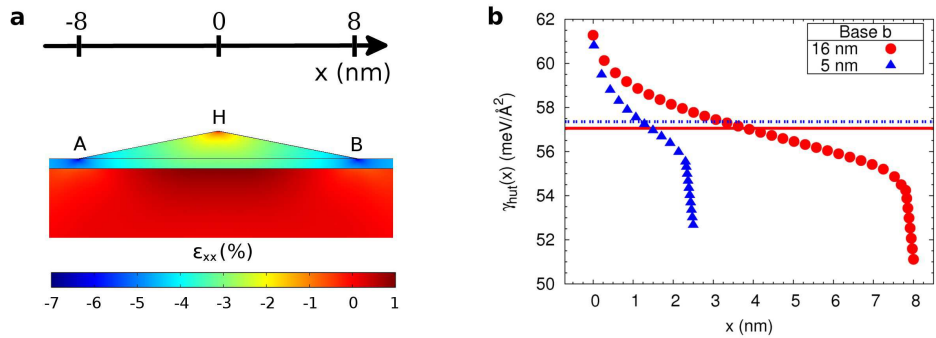

Fig. S5. (a) Map of the horizontal component of strain $\varepsilon_{xx}$ for the case with N=4 MLs and $b \sim b_{min}$ (16 nm) as obtained by FEM. The strain-dependent surface energy $\gamma_{hut}$ on every point of the surface is computed by merging this strain field with the dependence on the distance as given by Eq. (S2). The local values of the surface energy on the facet along HB are reported in panel (b) (red circles). A full red line reports its average along the facet, i.e. $\gamma_{hut}$ entering Eq. (S1) and (S7). For N=4, $\gamma_{hut}$ has a negligible dependence on the hut base, as shown by comparing the data for $b = 5$ nm (dashed blue line and blue triangles). This proves the validity of approximation leading to Eq. (S8). A larger sensitivity to the base is observed for smaller values of N (not shown).

As already stressed, the theoretical estimate for $b_{min}$ is very close to the experimentally observed one. Still, some of the assumptions on which our simple model is built suggest some care in trusting too much the quantitative agreement. For instance, the model directly considers mature, high aspect-ratio huts, avoiding dealing with the initial stages of formation where kinetics could play an important role, also in terms of formation of atomic-scale precursors [8]. Kinetic effects are not considered even in our high aspect-ratio limit, where full thermodynamic behavior is invoked in the minimization of Eq. (S7). Despite our effort to use realistic parameters, we recall that Γ was tentatively set based on a previous analysis on a similar system. Furthermore, we notice that DFT is not exact and it has not been clearly established to which extent, quantities like surface energies are reliable for the present system. Notice that by changing the value of $\Delta\gamma$ in Eq. (S8) by only ± 1 meV/Å$^2$, $b_{min}$ changes in the range (12–19) nm. Perhaps more importantly, all parameters were quantified by considering 100% Ge, whereas Si-Ge intermixing do take place during Ge deposition and the subsequent annealing process. While we cannot provide quantitative estimates, based on Eq. (S8) we can at least state that a 100%-Ge estimate of $b_{min}$ should underestimate the real value. If some Si atoms populate the top edge of the NW, for instance, then Γ should increase owed to the higher dangling-bond energy with respect to Ge. At the same time, if the NWs' facets host some silicon, then $\Delta\gamma$ should decrease, as reported in Ref. [4].

We believe that the importance of the model mainly resides in the prediction, per se, of the existence of a magic base. This nicely explains the extremely sharp distribution of wires' sizes. In turn, this result depends on the fact that exposing {105} facets provide energy lowering. Actually, for this system (and for the similar one investigated in Ref. [4]), surface-energy minimization is a much stronger driving force with respect to elastic-energy relaxation. For instance, for N=4 and $b=b_{min}$, the first term in the square brackets of Eq. (S7) is ~ 13 times smaller than the second (~ -0.045 meV/Å$^3$ and ~ -0.590 meV/ Å$^3$, respectively).

Finally, we wish to discuss, in general terms, the stability of wires with respect to real Stranski-Krastanow islands, such as dome-shaped islands [7]. Let us focus our attention on the stable wire, i.e. the one corresponding to $b_{min}$. For N=4, Eq. (S7) gives $(\Delta E/V)_{b=b_{min}} \equiv \rho_{min} = -0.345$ meV/Å$^3$. If the wire elongates, this quantity simply increases linearly. This is at variance with a typical self-similar evolution [7], where $\Delta E$ is given by the sum of a term linear in $V$ ($\Delta \rho_{eff}$), a term scaling as $V^{2/3}$ (accounting for surface costs), and one scaling as $V^{1/3}$ (edges). For sufficiently large volumes, then, $\Delta E \approx V \Delta \rho_{eff}$. As a dome island provides pronounced strain relaxation, $\Delta \rho_{eff}$ (<0) is much larger in absolute value with respect to the wire case. For instance, for a Ge dome, based on Ref. [2], we find $\Delta \rho_{eff} \approx -0.6$ meV/Å$^3$. This means that, starting from sufficiently large volumes, domes would eventually prevail even over the lowest-energy wire [9]. In order to predict a critical volume for energetic crossing between dome and wire shapes, one should quantify the unknown value of the dome edge energies. But even neglecting such terms (therefore overestimating domes' stability), we find that the magic-base wire should be more stable than a dome with equal volume up to a remarkable length of $\approx 0.6$ μm, corresponding to a dome base of 45 nm. It is clear that if a wire reaches such length, transformation to domes would be kinetically hindered, requiring a massive rearrangement of atoms, making it possible to observe much longer (metastable) wires.

**References**

[1] U. Denker, M. Stoffel, and O. G. Schmidt, Phys. Rev. Lett. **90**,196102 (2003).

[2] M. Brehm *et al*., Phys. Rev. B **80**, 205321 (2009).

[3] D. Scopece, F. Montalenti, and M. J. Beck, Phys. Rev. B **85**, 085312 (2012).

[4] G. Chen *et al*., Phys. Rev. Lett. **108**, 055503 (2012).

[5] A. Rastelli and H. von Känel, Surf. Sci. **515**, L493 (2002).

[6] F. Montalenti *et al*., Phys. Rev. Lett. **93**, 216102 (2004).

[7] F. Montalenti and L. Miglio, *"Silicon-germanium (SiGe) nanostructures: Production, properties and application in electronics"* (Ed. Y. Shiraki and N. Usami, Woodhead Publishing – UK, 2011).

[8] V. L. Arapkina and V. A. Yuryev, J. Appl. Phys. **109**, 104319 (2011).

[9] T. I. Kamins, G. Medeiros-Ribeiro, D. A. A. Ohlberg, and R. Stanley Williams, J. Appl. Phys. **85**, 1159-1171 (1999).